\documentclass[aps,preprintnumbers,superscriptaddress,nofootinbib,aps,prd,twocolumn,floatfix]{revtex4-1}
\pdfoutput=1

\usepackage{multirow,array}
\usepackage{amsmath,amssymb}
\usepackage{graphicx}
\usepackage{color}
\usepackage{url}
\usepackage{slashed}
\usepackage{multirow}
\usepackage{footmisc}
\usepackage{hyperref}
\usepackage{braket}
\usepackage{colortbl}

\hypersetup{
  pdfauthor={Dorival Goncalves, David Lopez-Val} 
  pdftitle={Boosting Alignment Limit  searches with $t\bar{t}A$},
  pdfkeywords={QCD, NLO, Higgs Physics}
}

\newcommand\one{\leavevmode\hbox{\small1\normalsize\kern-.33em1}}

\newcommand{\gev}{{\ensuremath\rm GeV}}
\newcommand{\tev}{{\ensuremath\rm TeV}}

\def\slashchar#1{\setbox0=\hbox{$#1$}           
   \dimen0=\wd0                                 
   \setbox1=\hbox{/} \dimen1=\wd1               
   \ifdim\dimen0>\dimen1                        
      \rlap{\hbox to \dimen0{\hfil/\hfil}}      
      #1                                        
   \else                                        
      \rlap{\hbox to \dimen1{\hfil$#1$\hfil}}   
      /                                         
   \fi}

\newcommand{\be}{\begin{eqnarray*}}
\newcommand{\ee}{\end{eqnarray*}}

\newcommand{\bee}{\begin{eqnarray}}
\newcommand{\eee}{\end{eqnarray}}
\newcommand{\beeq}{\begin{equation}}
\newcommand{\eeeq}{\end{equation}}

\newcommand{\PHiggs}{H}

 \newcommand{\sa}{\ensuremath{\sin\alpha}}
 \newcommand{\ca}{\ensuremath{\cos\alpha}}

 \renewcommand{\sb}{\ensuremath{\sin\beta}}

\newcommand{\unitarity}{Akeroyd:2000wc}

\newcommand{\superiso}{Mahmoudi:2007vz}
\newcommand{\higgsbounds}{Bechtle:2008jh}

\newcommand{\thdmcolliders}{Craig:2013hca}

\newcommand{\thdmnewphysics}{Carena:2002es}
\newcommand{\thdmpheno}{Gunion:2005ja}
 
\newcommand{\searchescpodd}{Schael:2006cr,CMS:2015mca,CMS:2012ywa,Aad:2014ioa,Aad:2014vgg,Khachatryan:2014wca,Aad:2014ioa,Khachatryan:2015qba,Abdallah:2004wy}
\newcommand{\searchescpeven}{Aad:2014ioa,Aad:2014vgg,Khachatryan:2014wca}
\newcommand{\searchescharged}{Abdallah:2003wd}

\newcommand{\ourchannel}{\ensuremath{t\bar{t}A(b\bar{b})}}

\begin{document}
\title{Boosting to identify: pseudoscalar searches with di-leptonic tops}
\author{Dorival Gon\c{c}alves} \email{dorival.goncalves@durham.ac.uk}
\affiliation{Institute for Particle Physics Phenomenology, Department
of Physics, Durham University, United Kingdom \\[0.1cm]}

\author{David L\'opez-Val} \email{david.lopezval@uclouvain.be }
\affiliation{Centre for Cosmology, Particle Physics \& Phenomenology CP3, Universit\'e catholique de Louvain,
Belgium}

\preprint{}

\begin{abstract}
 \noindent
The quest for new heavy states is a critical component of the LHC physics program.
In this letter, we study the search for pseudoscalar bosons produced in association with a $t\bar{t}$ pair.
We consider the final state $t\bar{t} A \to t\bar{t} b\bar{b}$ with di-leptonic top pair signature, 
and reconstruct the boosted $A \to b\bar{b}$ candidate with jet substructure techniques, achieving
a remarkable sensitivity over a broad range of pseudoscalar masses and Yukawa couplings. We apply
this strategy to a Type-I  Two-Higgs-Doublet Model, demonstrating its ability to probe a realistic, UV-complete
extended Higgs sector. In particular, we find that the $13~\tev$ LHC with $300~fb^{-1}$ of data can constrain
the region $\tan\beta>1.5$ at 95$\%$ CL for a light pseudoscalar with $m_A = 50$ GeV. Moreover, the whole
mass range $20~\gev<m_A<210~\gev$ can be ruled out for $\tan\beta \leq 1$. Finally, we show that it is also 
possible to directly probe the CP-structure of the heavy scalar, and hence to distinguish a CP-odd (A) from a 
CP-even (H) 2HDM resonance.
\end{abstract}

\maketitle

\tableofcontents

\section{Introduction}
\label{sec:intro}
Many of the fundamental questions brought to the fore
by the  Higgs-like 125 GeV discovery~\cite{discovery} remain as of today unanswered. 
A primordial one is  whether the observed resonance does in fact account for
electroweak symmetry breaking and mass generation in exactly the form
postulated by Higgs, Englert and Brout~\cite{higgs,Englert:1964et,Guralnik:1964eu}. 
Or if it is rather a first footstep into the Beyond: namely, the vast new physics
territory in which the 125~GeV particle would be part of an extended
Higgs sector, possibly within the reach of the LHC. Notwithstanding, the great agreement
between a pure SM-like Higgs hypothesis and the experimental data implies that, if actually 
around, additional Higgs partners can only mildly mix with the $125$~GeV state,
and thus have a likely unmeasurable impact on its properties~\cite{Corbett:2015ksa}. 

Direct scalar searches, for instance through heavy quark-rich final states~\cite{Khachatryan:2015tra,
CMS:2015yja,Craig:2015jba,Gori:2016zto,Hajer:2015gka,Craig:2016ygr,Kozaczuk:2015bea,
Casolino:2015cza}, are therefore a paramount avenue towards an extended Higgs sector. Among the
possible candidates, a relatively light pseudoscalar $A$ is not yet overly constrained, and may lie even in the few 
GeV range. Such mild bounds reflect in part that the $A$ state cannot couple at tree level to the SM vector bosons, 
but only through loop effects~\cite{dim6}. In addition, the overwhelming SM backgrounds render fermionic decays 
signatures troublesome to tackle. An emblematic case  is the process $pp(gg) \to A+$ jets  where, in spite of the 
possibly large rates, the $A\rightarrow b\bar{b}$ mode is experimentally inaccessible. One way around the challenging 
QCD environment  has been identified, e.g., for the SM Higgs associated production along with leptons in the channels 
$Zh$~\cite{Butterworth:2008iy} and $tth$~\cite{Plehn:2009rk,Buckley:2015vsa}. With the use of jet substructure techniques,  
both channels can help to access the dominant, yet challenging, decay mode $h\rightarrow b\bar{b}$.  Likewise,  heavy 
quark-rich final states could be operative to pin down pseudoscalar states through $A\rightarrow b\bar{b}$, and also 
$t\bar{t}A$ with fully leptonic top-quark decays. 

In addition to setting bounds on the pseudoscalar mass-coupling strength plane, these decay modes also grant direct 
access to the  CP-structure of the new state through the spin correlations analysis~\cite{Buckley:2015vsa}. A direct separation 
between the CP-even ($H$) from the CP-odd ($A$) hypotheses would be a primary task following an eventual signal excess. As a
matter of fact, typical multi-Higgs extensions, such as the Two-Higgs-Doublet Model (2HDM), include both types of bosons, with 
rather similar collider footprints in the scenarios best compatible with the current data. \medskip

Our aim in this letter is to cover a wide range of masses up to $m_A\sim2m_t$ in 
a pseudoscalar search through the usually dominant channel $A\rightarrow b\bar{b}$,
where the CP-odd state is produced through the $t\bar{t}A$ mode with di-leptonic tops, and
jet substructure techniques are exploited to reconstruct the resonance candidate. 
Similarly to Refs.~\cite{Kozaczuk:2015bea,Casolino:2015cza}, our starting logics is to simulate our signal with a 
Simplified Model. We show the power of this analysis to significantly constrain the parameter space
of a Simplified Model, by adapting the jet substructure tagging for different mass regimes. 
We find competitive sensitivities, most significantly for pseudoscalar mass and Yukawa coupling 
ranges which are arduous to access otherwise. Next, we examine the implications 
of this search on the \emph{alignment without decoupling} limit of the 2HDM. So doing, 
we assess the ability of the $\ourchannel$ analysis to probe realistic scenarios of a UV-complete 
extended Higgs sector. Lastly, we show that this channel also grants direct access to the CP-structure 
of such a hypothetical novel resonance.
\medskip

This paper is organised as follows. In Section~\ref{sec:ana} we provide the details of the signal and 
background simulation and the boosted $A\rightarrow b\bar{b}$ reconstruction, illustrating the results
of our collider analysis in the Simplified Model setup. The corresponding interpretation in terms of the 2HDM
is discussed in Section~\ref{sec:interpretation}. Section~\ref{sec:cp} focuses on the direct
CP-measurement of the heavy scalar resonance through  spin correlations.
Finally, a summary of our key findings is delivered in Section~\ref{sec:summary}. 

\section{Analysis}
\label{sec:ana}
\begin{table*}[t!]
\centering
\begin{tabular}{l | c  | c | c 
|| c  | c | c
|| c  | c | c }
  \multicolumn{1}{c}{} &
  \multicolumn{3}{c ||}{C/A $R=0.6$} &
  \multicolumn{3}{c ||}{C/A $R=1.2$} &
  \multicolumn{3}{c }{C/A $R=2.4$} 
    \\
 \cline{2-10}
  \multicolumn{1}{c}{} &
  \multicolumn{1}{c|}{$t\bar{t}A_{50}$}&
  \multicolumn{1}{c|}{$t\bar{t}b\bar{b}$} &
  \multicolumn{1}{c||}{$t\bar{t}Z$} &
  \multicolumn{1}{c|}{$t\bar{t}A_{150}$}&
  \multicolumn{1}{c|}{$t\bar{t}b\bar{b}$} &
  \multicolumn{1}{c||}{$t\bar{t}Z$} &
  \multicolumn{1}{c|}{$t\bar{t}A_{200}$}&
  \multicolumn{1}{c|}{$t\bar{t}b\bar{b}$} &
  \multicolumn{1}{c}{$t\bar{t}Z$} \\
  \hline
{BDRS $A$-tag,  $p_{T\ell}>15$~GeV,  $|\eta_\ell|<2.5$}   
&\multirow{2}{*}{ 1.08}  & \multirow{2}{*}{7.95} & \multirow{2}{*}{0.99}  
&\multirow{2}{*}{ 0.63}  & \multirow{2}{*}{10.93} & \multirow{2}{*}{1.11} 
&\multirow{2}{*}{ 0.52}  & \multirow{2}{*}{11.69} & \multirow{2}{*}{1.29}    \\ 
$n_\ell = 2$, $p_{Tj}>30$~GeV,  $|\eta_j|<2.5$, $n_j\ge 2$ 
&&&
&&&
&&&\\
two extra b-tags -- four in total  
& 0.43  &  3.77 & 0.19
& 0.26  &  4.21 & 0.21
&0.14   &  3.11 & 0.14
\\ 
$|m_{A}^{\text{BDRS}}-m_A|/m_A<0.15$
& 0.22   & 0.35  & --
& 0.085 & 0.26  & --
& 0.05   &  0.35  & -- \\ 
\hline
$m_{ll}>75$~GeV
& 0.15  & 0.10  & --
& 0.06  & 0.21& --
& 0.04 & 0.20 &  --
\end{tabular} 
\caption{Signal and background cut-flow analysis for the LHC at $\sqrt{s}=13$~TeV. Rates are shown  in fb 
accounting for b-tagging efficiencies, hadronisation and underlying event effects. For illustrative purposes,
we display the results for three different signal hypotheses $m_A=50,150,200$~GeV calculated
respectively with C/A $R=0.6,1.2,2.4$. The branching ratios $\mathcal{BR}(A\rightarrow b\bar{b})$ are
calculated with \textsc{Hdecay} is included and we assume $\tan\beta=1$. All signal and backgrounds samples 
are produced at NLO precision with the \textsc{MC@NLO} algorithm.} 
\label{tab:cuts}
\end{table*}

We study the signature of a pseudoscalar $J^\pi = 0^-$ state produced in association with a top pair $t\bar{t}A$ at 
the 13~TeV LHC. We access this channel via the decay $A\rightarrow b\bar{b}$ along with di-leptonic tops. To simulate
the signal, we resort to a Simplified Model which describes the dynamics of the $0^-$ state through the interaction 
Lagrangian 
\begin{alignat}{5}
  \mathcal{L}\supset 
  \kappa_t\frac{iy_{t}A\bar{t}\gamma_5 t}{\sqrt{2}}
  +\kappa_b\frac{iy_{b}A\bar{b}\gamma_5 b}{\sqrt{2}} \, ,
  \label{eq:L_simp}
\end{alignat}
where $y_{t(b)} \equiv \sqrt{2}m_{t(b)}/v$ are the SM Yukawa couplings to top (bottom) quarks,
with the Higgs {\it vev} $\braket{H} = v/\sqrt{2} \simeq 246$ GeV. This is a four-parameter model, 
where the input quantities are chosen to be: the pseudoscalar mass $m_A$, its width $\Gamma_A$, and
the top (bottom) rescaling factors $\kappa_{t(b)}$, through which we can test the relative strength of their 
respective Yukawas to the new $0^-$ resonance. By keeping $\Gamma_A$ as a free parameter, we allow
our framework to accommodate additional pseudoscalar couplings to other new degrees of freedom. As we 
will discuss in more detail in Section~\ref{sec:interpretation}, such a generic Simplified Model setup dovetails 
with a broad class of more specific new physics models such as extended Higgs sectors like the 2HDM, or Dark 
Matter (DM) models~\cite{DM1,Goncalves:2016bkl,Harris:2014hga,Haisch:2012kf,Abdallah:2015ter}.  For the latter, 
the Lagrangian in Eq.~\ref{eq:L_simp} can be extended with the extra pseudoscalar mediator coupled also to the  DM 
particle.  This way, the strategy we explore in this paper can be seen as the counterpart  of a DM search where, instead
of probing the mediator decays to the Dark Sector,  we now test the interactions of the pseudoscalar  bouncing back to
the SM. \medskip

\begin{figure*}[htb]
  \includegraphics[width=.33\textwidth]{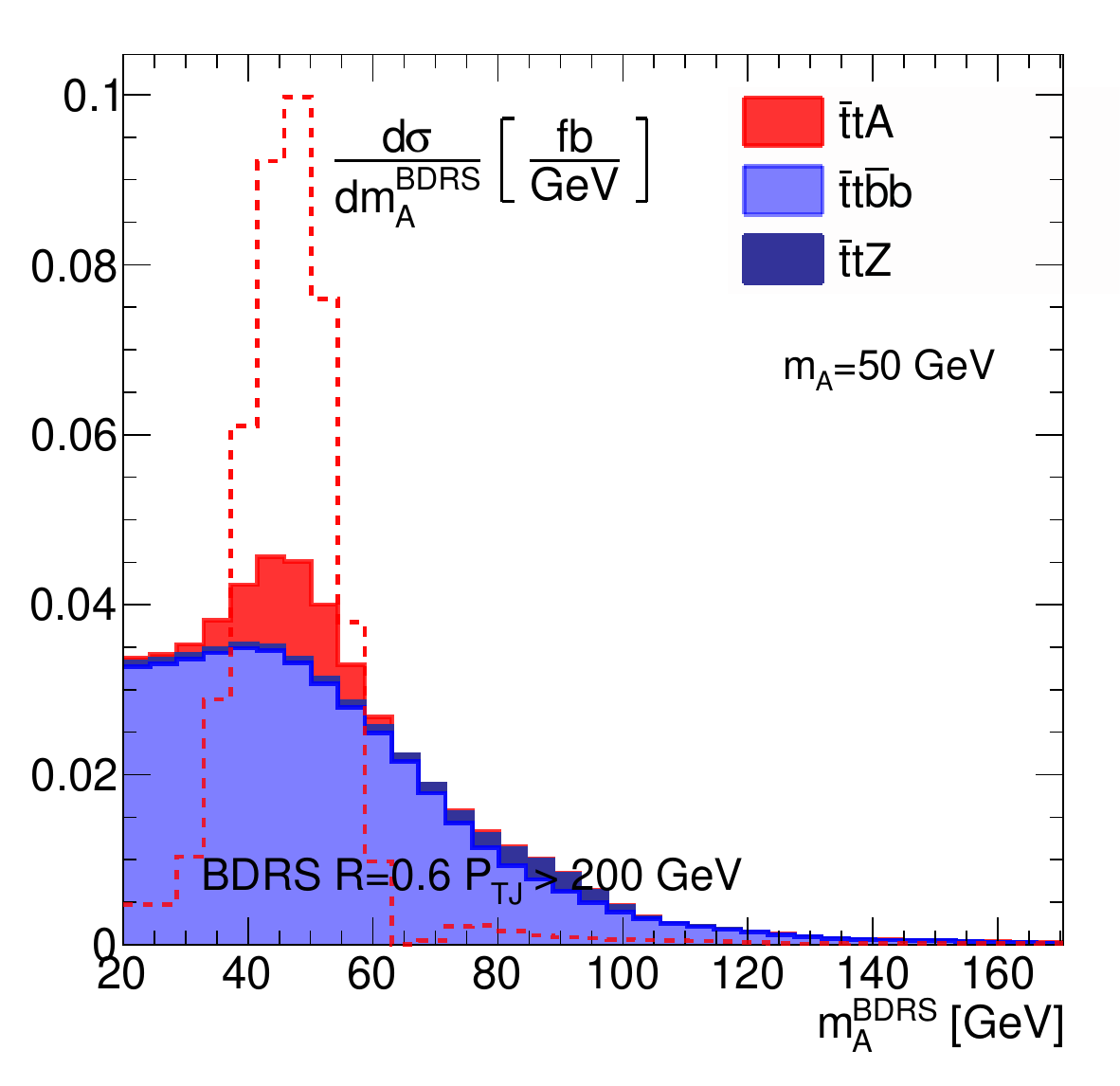}
  \hspace{-0.2cm}
  \includegraphics[width=.33\textwidth]{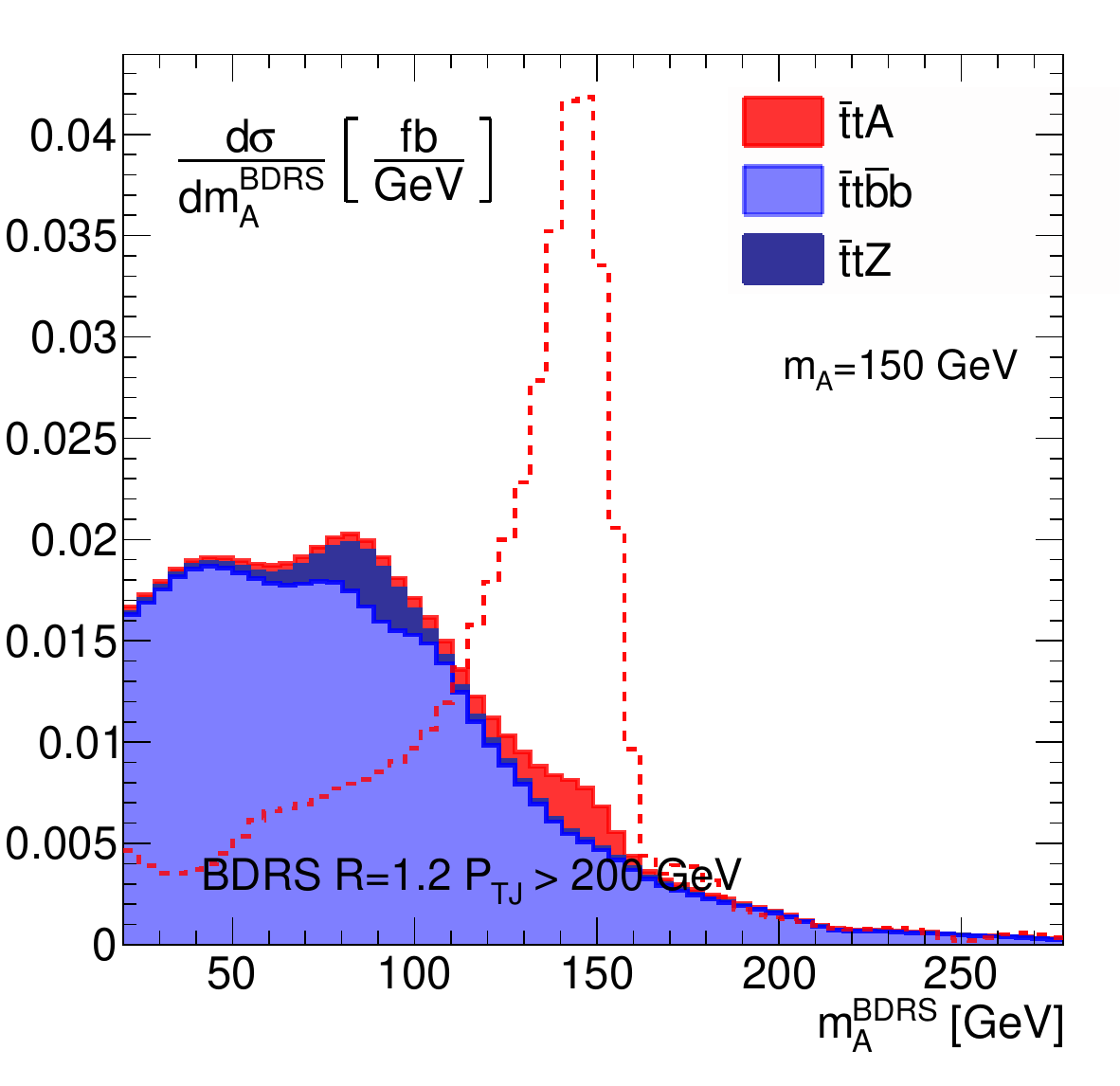}
    \hspace{-0.2cm}
  \includegraphics[width=.33\textwidth]{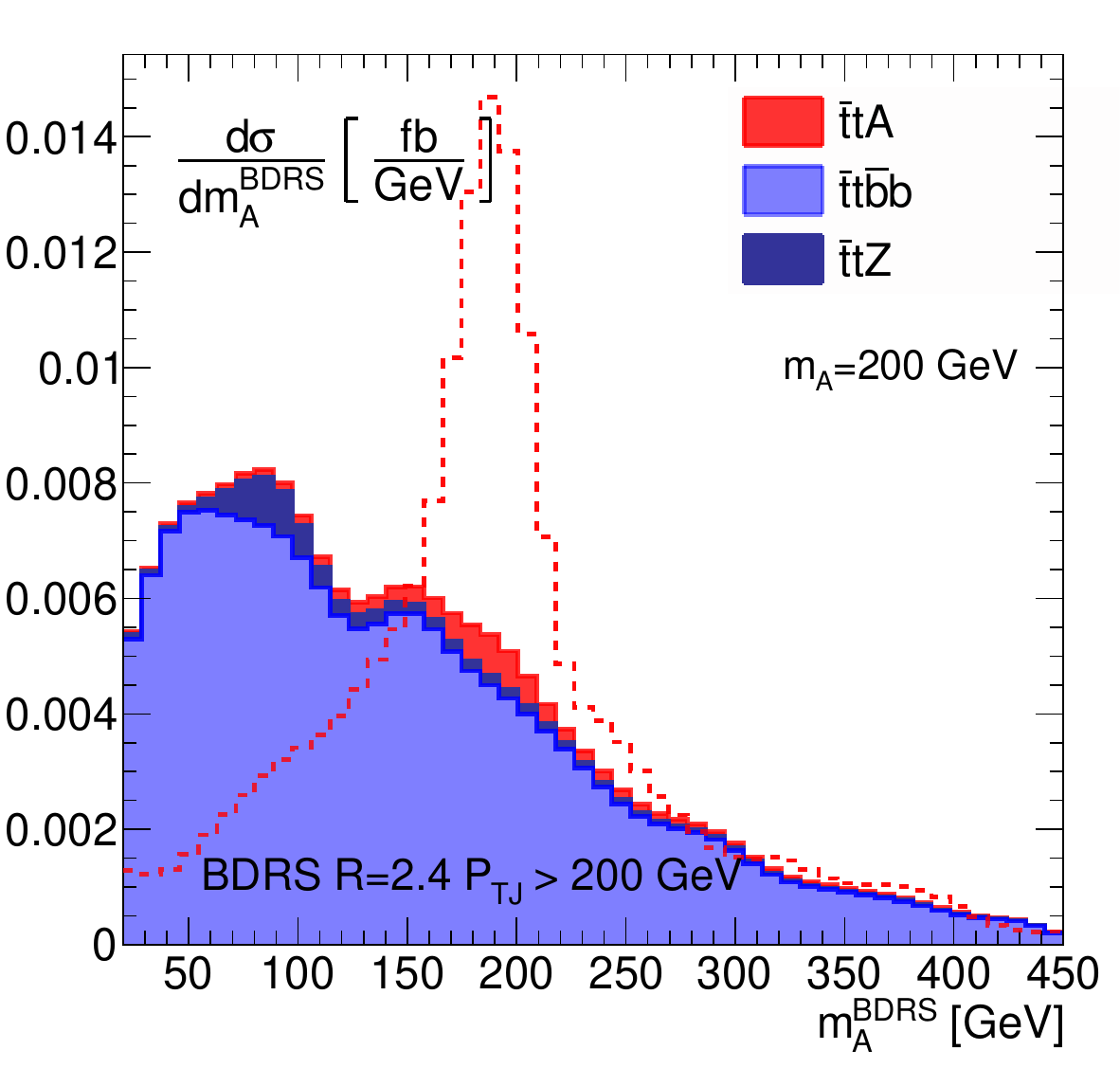}
  \caption{Signal and background invariant mass distribution for the BDRS tagged fat-jet $m_A^{BDRS}$. We show the three different signal
   hypotheses presented on Tab.~\ref{tab:cuts}: $m_A=50,150,200$~GeV calculated respectively with C/A $R=0.6,1.2,2.4$.
  The histograms are stacked. We also display the normalised signal component to the total background rate (red dashed).}
  \label{fig:mj}
\end{figure*}
\begin{figure}[t!]
\vspace{-0.5cm}
  \includegraphics[width=.47\textwidth]{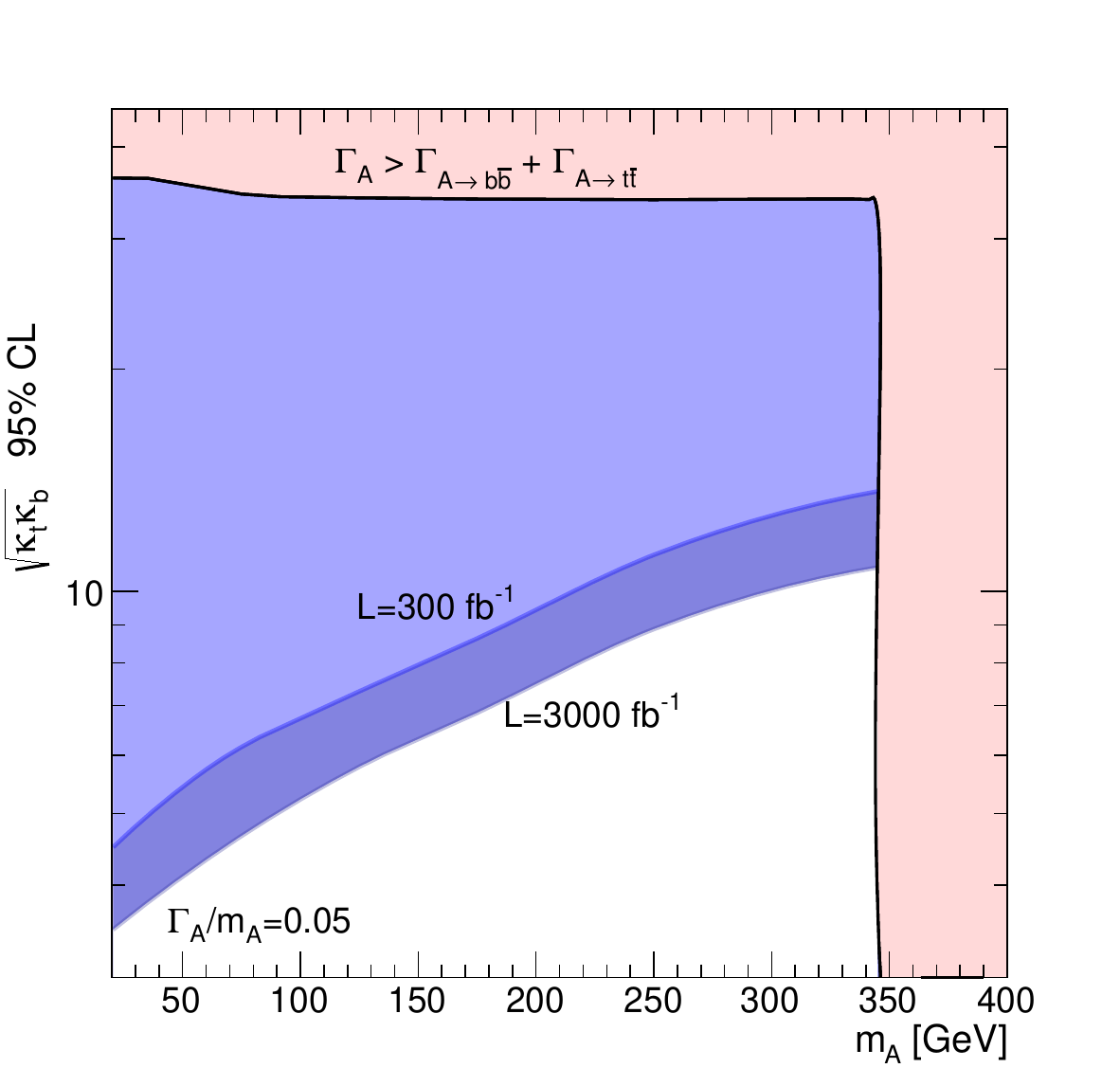}
  \vspace{-0.5cm}
  \caption{95\% CL exclusion region on $\sqrt{\kappa_t\kappa_b}$ as a function of the  pseudoscalar mass $m_A$ in the Simplified Model
  framework Eq.~\ref{eq:L_simp}. The shaded regions in blue can be excluded by the 13~TeV LHC with ${\mathcal{L}=300~fb^{-1}}$ (light blue) and
  ${3000~fb^{-1}}$ (dark blue). The shaded region in red cannot be probed and is given by the consistency constraint 
   $\Gamma_A\ge \Gamma_{A\rightarrow t\bar{t}} + \Gamma_{A\rightarrow b\bar{b}}$. We assume $\Gamma_A/m_A=0.05$.}
  \label{fig:cls_simp}
\end{figure}

To control the backgrounds, we require four b-tagged jets. The major backgrounds  are  $t\bar{t} b \bar{b}$ and $t\bar{t}Z$. 
The $t\bar{t}A$ signal sample is generated with \textsc{MadGraph5+Pythia8}~\cite{mg5,pythia8}, while for
the backgrounds $t\bar{t}b\bar{b}$ and $t\bar{t}Z$ we use \textsc{Sherpa+OpenLoops}~\cite{sherpa,openloops}.
A proper modelling for the QCD effects is of major importance in this study as the Higgs is part of a multi-jet system.
Hence, we simulate all samples at Next-to-Leading Order (NLO) QCD with the \textsc{MC@NLO} algorithm~\cite{mcatnlo}. 
To evaluate the signal decay rates  $A\rightarrow b\bar{b}$, we use the \textsc{2HDM} predictions from the \textsc{Hdecay}~\cite{hdecay}
package,  with appropriately rescaled Yukawas via the parameter $\tan\beta$ (cf. Section~\ref{sec:interpretation} further
down). Additional non-SM effects from higher orders are precluded by setting the \textit{alignment limit} condition $\cos(\beta-\alpha)=0$. 
Spin correlation effects in the top decays are fully accounted for in our simulation~\cite{madspin,sherpa_spin}.  We also consider 
the hadronisation and underlying  event effects with the ~\textsc{Pythia8} and \textsc{Sherpa} modules. \medskip

We start our analysis by requiring two isolated opposite sign leptons with $p_{T\ell}>15$~GeV and $|\eta_\ell |<2.5$. The leptons
are defined as isolated if there is less than 20\% of hadronic activity around the leptonic radius ${R=0.2}$.  For the hadronic 
part of the event, we start by reclustering jets with the Cambridge/Aachen (C/A) jet-algorithm with the \textsc{Fastjet} package~\cite{fastjet}. 

The opening angle between the two $b$-quarks generated from the pseudoscalar decay can be estimated in the boosted regime by 
\begin{alignat}{5}
\Delta R_{b\bar{b}}\sim \frac{2m_A}{p_{TA}} ,
  \label{eq:Rbb} 
\end{alignat}
where $p_{TA}$ stands for the pseudoscalar transverse momentum. 
As we probe a wide range of pseudoscalar masses, we customise our search with different jet radius $R$ for distinct 
pseudoscalar mass range hypotheses: $R=0.6,1.2$ and 2.4 for $m_A=[20,100)$, $[100,200]$~GeV and (200,400]~GeV, 
respectively. We require at least one fat jet with $p_{TJ}>200$~GeV and ${|\eta_J|<2.5}$. This transverse momentum  
selection on the fat-jet is enhanced to $p_{TJ}>310$~GeV for very large pseudoscalar masses $m_A=(300,400]$~GeV. 
Instead  of requiring a larger jet radius $R>2.4$, which would also collect undesired radiation from the top decays, we 
find more efficient to demand a  larger transverse momentum selection in this regime.  

The fat-jet is  demanded to be tagged by the BDRS algorithm~\cite{Butterworth:2008iy}. The BDRS  \emph{filtering} 
promotes the  invariant  fat-jet mass to be a robust  observable as it efficiently  controls the pile-up effects~\cite{ATLAS:2012am}.  The
two hardest sub-jets from this tagged jet are then $b$-tagged.  We adopt 70\% $b$-tagging efficiency and 1\% misstag 
rate~\cite{btagging}.\medskip

For the remaining hadronic activity, we remove the tagged fat-jet and recluster jets again with C/A but now 
with $R=0.5$, ${p_{Tj}>30}$~GeV and $|\eta_j|<2.5$. The smaller jet radius suppresses the possible underlying event contamination.
We require two extra b-tagged jets to suppress the possible extra backgrounds. Lastly, we demand the filtered fat-jet  mass $m_A^{BDRS}$
to  be in a window centered on the pseudoscalar mass $m_A$, ${|m_A^{BDRS}-m_A|/m_A<0.15}$. The cut-flow analysis is displayed in 
Tab.~\ref{tab:cuts}. For illustrative purposes, we show the results for three different signal hypotheses ${m_A=50,150,200}$~GeV with
$\tan\beta=1$. 

The signal and backgrounds invariant mass distributions $m_A^{BDRS}$ are shown in Fig.~\ref{fig:mj}. We display the three different
signal hypotheses presented on Tab.~\ref{tab:cuts}. The results include event selections up to the four b-jet tagging. The signal displays a clear 
peak structure  around the resonance mass, $m_A$, for all considered cases. We emphasise that this was only achievable after tailoring the jet 
substructure analysis in different mass regions through convenient fat-jet radius $R$ and transverse momentum selection $p_{TJ}$, as previously
described.\medskip

Within our Simplified Model framework, the  signal cross section -- or more generally any signal distribution -- can be written as
${\sigma_{sig}=\kappa_t^2\kappa_b^2\sigma(m_A,\Gamma_A)}$. Therefore, the analysis depends only on three parameters: 
the pseudoscalar mass $m_A$, its width $\Gamma_A$, and the product of the rescaling couplings to tops and bottoms $\kappa_t\kappa_b$. 
These parameters are only bound  \emph{a priori} to the consistency condition 
\begin{alignat}{5}
\Gamma_A\ge \Gamma_{A\rightarrow t\bar{t}} + \Gamma_{A\rightarrow b\bar{b}}
 \label{eq:consistency}\; ,\end{alignat}
which reflects that the pseudoscalar state is likely part of a larger UV completion, with possibly additional decay modes
besides the explicit ones in our Simplified Model. For simplicity,  we fix $\Gamma_A=0.05m_A$ throughout our Simplified Model analysis. \medskip

Based on the presented collider study, we can derive constraints on the Simplified Model parameters by
further exploring the kinematics of the leptonic top decays via a two-dimensional binned log-likelihood analysis on $(\Delta\eta_{ll},\Delta\phi_{ll})$. 
See Sec.~\ref{sec:cp} for a detailed discussion on the  $\Delta\phi_{ll}$ sensitivity.
In Fig.~\ref{fig:cls_simp} we display the 95\% CL limits on $\sqrt{\kappa_t\kappa_b}$ as a function of the pseudoscalar mass $m_A$. 
The accessible regions are limited  from above and below. The upper bound originates from the consistency condition 
Eq.~\eqref{eq:consistency}.  For all points inside the red area, the partial pseudoscalar widths to tops and bottoms would surpass 
the assumed total width,  $\Gamma_A=0.05m_A$. The lower bound stems from the limited statistics, as for all points below
the light (dark) blue bands would yield unobservable signal rates at the 13 TeV LHC after $300$ (3000)~fb$^{-1}$ 
of integrated luminosity.\medskip 

\section{A BSM scenario}
\label{sec:interpretation}

\subsection{Extended Higgs sector} 
Thus far we have applied our $\ourchannel$ analysis to the test-ground Simplified Model laid out 
by the Lagrangian~\eqref{eq:L_simp}. 
By trading completeness in favor of generality and a minimal 
number of  new parameters and fields, Simplified Models of this guise have become a cherished toolkit 
to analyse plausible new physics signal topologies and to interpret the results of collider searches~\cite{Alves:2011wf}. 
In the following, we promote our search strategy 
to a more realistic extended Higgs sector.
Our prime candidate is the \textsc{2HDM}~\cite{Gunion:1989we,Branco:2011iw}. 
In addition to its indisputable interest for Higgs coupling analyses~\cite{fits,Bernon:2015qea,Bernon:2015wef},  
collider searches ~\cite{\thdmcolliders}, and 
beyond the SM phenomenology~\cite{\thdmpheno}, the \textsc{2HDM} 
describes the low-energy Higgs sector of a variety of TeV-scale new physics models~\cite{\thdmnewphysics}, and 
contains all the necessary
ingredients to reinterpret the above Simplified Model analysis:
\begin{enumerate}
 \item{A CP-odd scalar $A$, which is present in the 2HDM physical spectrum, along with
 two neutral CP-even scalars $h,H$ and a pair of charged scalars $\PHiggs^\pm$;}
\item{A limit of \emph{alignment without decoupling}~\cite{Carena:2013ooa,Delgado:2013zfa}, 
in which one of the neutral CP-even mass eigenstates mimics the SM Higgs boson,
while at least one of the extra resonances remains relatively light. Precisely due to the
excellent agreement with the LHC data, the \emph{alignment limit} is strongly favored by the global LHC
fits~\cite{Belanger:2013xza} -  and at the same time very difficult to probe,
unless the additional states are not decoupled, and hence possible to pin down at colliders. 
This is therefore the natural scenario our strategy suits best.}
 \item{A minimal, UV complete embedding for fully flexible couplings~\cite{Lopez-Val:2013yba},
which in particular allow for both enhanced and suppressed fermion Yukawas.
This is possible in the \textsc{2HDM} as it includes two weak doublets which can couple to fermions
and gauge bosons independently, unlike other models such as the singlet extension~\cite{singletmodel}.}
\end{enumerate}

At variance with the couplings to the weak bosons, 
the Higgs-fermion Yukawas are not uniquely determined
by the underlying gauge structure. 
The four canonical \textsc{2HDM} setups ~\cite{Branco:2011iw} are obtained
by imposing a global $\mathcal{Z}_2$ invariance $\Phi_{1,2} \to \mp \Phi_{1,2}$
and linking each fermion
family to only one of the Higgs doublets $\Phi_i$. In turn, these are particular
cases of the more general \emph{Minimal Flavor Violation} category~\cite{mfv},
which include among others
the so-called \emph{flavor-aligned} 
\textsc{2HDM}~\cite{Pich:2009sp}. The Yukawa structures in the latter case can be parametrised through independently 
variable angles (cf. Tab.~\ref{tab:yukawas}) and may be seen as an interpolation 
between the canonical Type-I ($\gamma_b = \pi/2$) and Type-II models ($\gamma_b = 0$).

\begin{table}[t]
\begin{center} \begin{small}
\renewcommand{\arraystretch}{2.0}
\begin{tabular}{l|c|c|c} \hline
 & \multicolumn{1}{c|}{$h$} & \multicolumn{1}{c|}{$H$}& \multicolumn{1}{c}{$A$} \\ \hline
$\kappa_t $ & $\dfrac{\ca}{\sb}$ &  $\dfrac{\sa}{\sb}$ & $\cot\beta$ \\
$\kappa_b $ & $-\dfrac{\sin(\alpha-\gamma_b)}{\cos(\beta-\gamma_b)}$ & $\dfrac{\cos(\alpha-\gamma_b)}{\cos(\beta-\gamma_b)}$ & $\tan(\beta-\gamma_b)$ \\
\hline 
\end{tabular}
\end{small} \end{center}
\caption{Neutral Higgs boson couplings to fermions in a generic 2HDM, where the Yukawa interactions follow
a flavor alignment pattern parametrised through the independent angle $\gamma_{b}$ in the notation of~\cite{Bai:2012ex}.}
\label{tab:yukawas}
\end{table}

For a given choice of $\gamma_b$,  our $\ourchannel$-based strategy directly tests the 2-dimensional slice $(m_A,\tan\beta)$. The parameter 
$\tan\beta$  denotes as customary the {\it vev} ratio $\tan\beta \equiv \braket{\Phi_2}/\braket{\Phi_1} = v_2/v_1$ of the  individual Higgs fields.
In the \emph{alignment limit},  the second mixing angle $\alpha$, which parametrises the rotation
of the neutral CP-even gauge-eigenstates into the physical fields $h,H$, is fixed through either $\cos(\beta-\alpha)=0$ or $\sin(\beta-\alpha )= 0$.
Mapping these parameters back onto the Simplified Model Lagrangian~\eqref{eq:L_simp}, we find
\begin{alignat}{5}
 \kappa_t = \cot\beta \;; \quad   \kappa_b = \tan(\beta-\gamma_b) = \cfrac{\tan\beta -\tan\gamma_b}{1+ \tan\beta\tan\gamma_b}
 \label{eq:map}\, .
\end{alignat}

\begin{figure}[!t]
  \includegraphics[width=.4\textwidth]{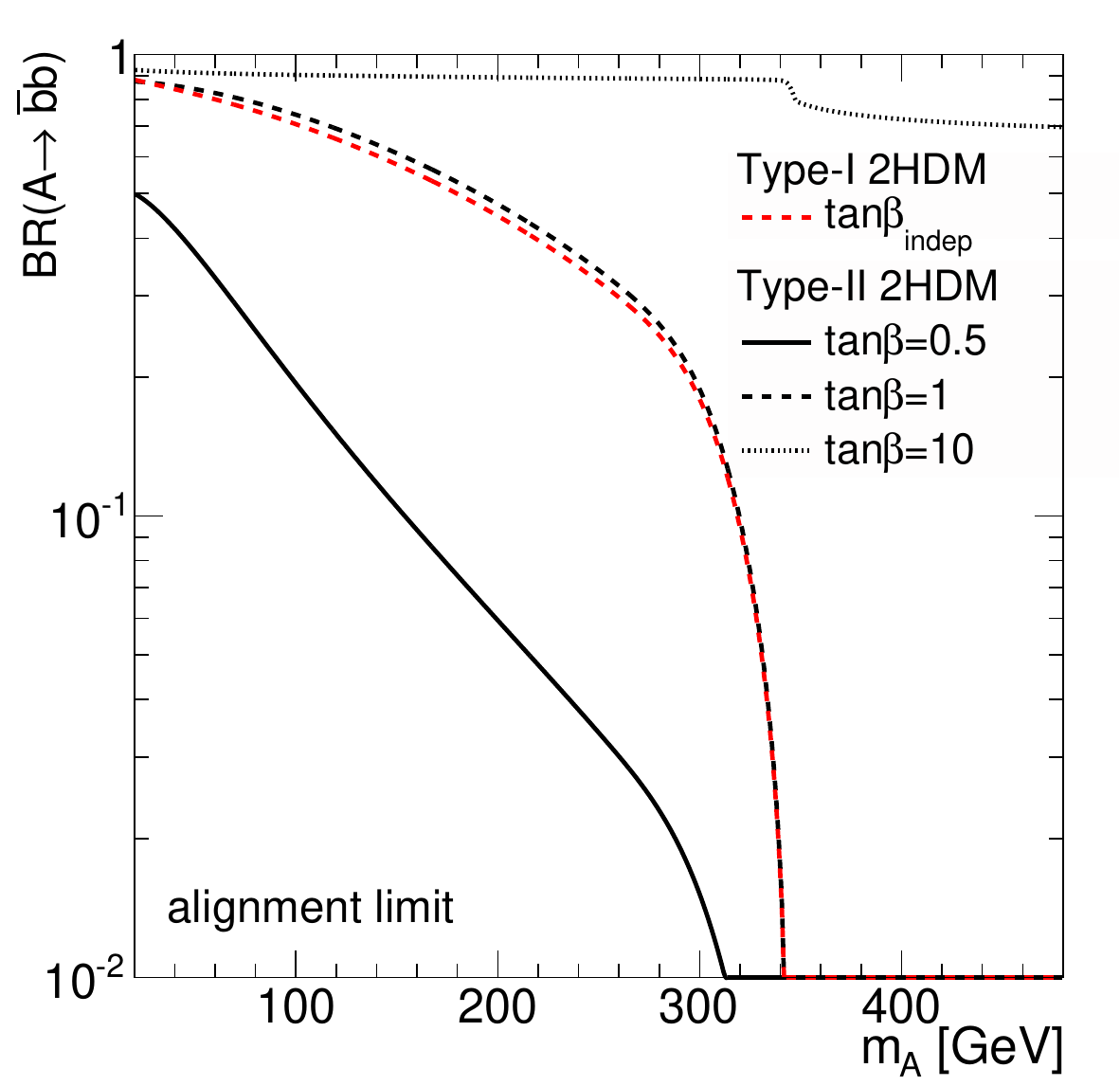}
  \caption{Branching ratio $\mathcal{BR}(A\rightarrow \bar{b}b)$ as a function of the pseudoscalar
  mass $m_A$ for Type-I  (red) and Type-II  (black) \textsc{2HDM} for exemplary scenarios in the \emph{alignment limit}.}
  \label{fig:branch}
\end{figure}
Accordingly, the exclusion contours for $\sqrt{\kappa_t \kappa_b}$ in Fig.~\ref{fig:cls_simp} impose constraints 
on $\tan\beta$ for every given $\gamma_b$ and  pseudoscalar mass hypothesis $m_A$. Eq.~\eqref{eq:map} makes it patent 
that the sensitivity attainable by the $\ourchannel$ analysis becomes optimal for a Type-I setup and slight variations thereof 
($\gamma_b \simeq \frac{\pi}{2}$), where both the top and the bottom Yukawas increase simultaneously for 
$\tan\beta < 1$ as $\sqrt{\kappa_t\kappa_b} = 1/\tan\beta$, generating a quadratic signal strength dependence 
$\mu_{\text{sig}} \propto \cot^2\beta$. In contrast, this search is loosely efficient for Type-II, where the top and the bottom 
Yukawas have inverse scalings with respect to each other. Here, a growth in the $\bar{t}tA$ production rate is counterbalanced
by a suppressed $A \to bb$ decay. The dependence on $\tan\beta$ is marginal here, and features merely through the total 
decay width $\Gamma_A$. 

To better visualize the key parameter dependences, in Fig.~\ref{fig:branch} we display the pseudoscalar branching ratio 
$\mathcal{BR}(A\rightarrow b\bar{b})$ as  a function of its mass  $m_A$. All decay rates are computed with \textsc{Hdecay}~\cite{hdecay}
for the \textsc{2HDM} in the \emph{alignment limit}. For definiteness we assume no additional chain decays 
$A \to hZ, HZ, H^{\pm} W^{\pm}$. For the Type-II model  we choose as illustrative {\it vev} ratios ${\tan\beta=0.1,1,30}$, while for Type-I the 
results do not depend on $\tan\beta$. We see how ${\mathcal{BR}(A\rightarrow b\bar{b})}$ is enhanced (suppressed) for large (small) 
values of $\tan\beta$ in the Type-II  \textsc{2HDM}, reflecting the different rescalings of the top and the bottom
Yukawas. The steep fall around $m_A\gtrsim 2m_t$ signals the opening of the top pair mode 
$A\rightarrow t\bar{t}$.  In a Type-II model with sufficiently large $\tan\beta$, the enhanced bottom Yukawa explains why $A \to b\bar{b}$ 
dominates even  above the top pair threshold.  The mild, yet visible, offset between the $\mathcal{BR}(A\rightarrow b\bar{b})$ curves 
for the Type-I  with respect to the Type-II \textsc{2HDM} at $\tan\beta \simeq 1$ follows from the different relative signs between the
top and the bottom Yukawas, which feature through the top and bottom loop interferences contributing to the loop-induced 
modes $A\rightarrow gg,\gamma\gamma$.

\subsection{Parameter space}

In the following, we illustrate how our proposed search strategy is capable to constrain phenomenologically viable 
\textsc{2HDM} scenarios in the well-motivated \emph{alignment without decoupling} limit.
For that, in Tab.~\ref{tab:interpretation}, we identify exemplary benchmarks
spanning the $m_A - \tan\beta$ plane tested by our search. For definiteness, we stick hereafter to quark Yukawa couplings of Type-I. 

\begin{table}[t!]
\begin{center}
\begin{tabular}{|ll|}
\hline
\multicolumn{2}{|l|}{\cellcolor[gray]{0.9} \qquad \textbf{BI: $m_A > 125 $} \quad $\cos(\beta-\alpha) =0$ }\\ \hline\hline
$\bullet$\, $m_h = 125$ & \quad  $\bullet$\, $m_H = (220-300)$  \\ 
$\bullet$\, $m_{H^\pm} = \text{max}(175,m_A)$ & \quad $\bullet$\, $ m^2_{12} = \cfrac{m^2_A \tan\beta}{1+\tan^2\beta}$  \\ \hline\hline 
\multicolumn{2}{|l|}{\cellcolor[gray]{0.9} \qquad \textbf{BII: $63 < m_A < 125$} \quad $\sin(\beta-\alpha) =0$ }\\ \hline
$\bullet$\, $m_h = 120$ & \quad $\bullet$\, $m_H = 125$  \\ 
$\bullet$\, $m_{H^\pm} = 175$ & \quad  $\bullet$\, $ m^2_{12} = 0$   \\ \hline \hline
\multicolumn{2}{|l|}{\cellcolor[gray]{0.9} \qquad \textbf{BIII: $m_A < 63$}  \quad $\sin(\beta-\alpha) =0$ }\\ \hline
$\bullet$\, $m_h = 120$ & \quad $\bullet$\, $m_H = 125$  \\ 
$\bullet$\, $m_{H^\pm} = 175$ & \quad  $\bullet$\, $ m^2_{12} = \cfrac{(m_H^2 + 2m_A^2)\tan\beta}{2(1+\tan^2\beta)}$  \\ \hline \hline
\end{tabular}
\end{center}
 \caption{Sample benchmarks for a Type-I \textsc{2HDM}   
 in the different patches of the $m_A - \tan\beta$ plane covered by the $\ourchannel$ search. All masses are given in GeV.}
 \label{tab:interpretation} 
\end{table}
We separately cover the two relevant pseudoscalar mass ranges, namely above (below) the SM Higgs mass.
For $m_A > 125$ GeV (resp. $m_A < 125$ GeV) we assume a \textit{direct} (\text{flipped}) CP-even eigenmass ordering,
fixing the SM-like Higgs mass to $m_h = 125$~GeV 
($m_H = 125$~GeV) and $\beta-\alpha$ through the appropriate \emph{alignment} condition. The bosonic
chain decays ${A\to hZ, HZ, H^{\pm} W^{\pm}}$ do not contribute in any case.
Compatibility with all model constraints we assess through an in-house interface of the public tools { \sc 2HDMC}~\cite{Eriksson:2009ws}, 
{ \sc HiggsBounds}~\cite{\higgsbounds},  {\sc SuperIso}~\cite{\superiso} and {\sc{HiggsSignals}}~\cite{Bechtle:2013xfa} along with additional 
own routines (cf. also ~\cite{Kling:2016opi} for an up-to-date review). The average LHC signal strength of the SM Higgs boson is by construction
satisfied in the \emph{alignment limit}, with the proviso that non-standard Higgs decays are suppressed or simply closed. 
For that, in the BIII region we fix the $Z_2$ soft-breaking mass $m^2_{12}$~\cite{Branco:2011iw} such that the
mode $H \to AA$ vanishes at tree-level. The relatively small mass splittings between the different additional Higgs states
agree with Electroweak Precision Observables~\cite{ewpo,Baak:2014ora}.
The lines $m_{12}^2 \simeq  m_A^2 \tan\beta / (1+\tan^2\beta)$ and
$m^2_{12} = 0$ for the BI (resp. BII) regions satisfy 
unitarity~\cite{\unitarity,Gunion:2002zf}, perturbativity~\cite{Chen:2013kt} and vacuum stability~\cite{vacuum}.
We test values of $\tan\beta$ down to $\sim 0.1$, assuming that these already border
the onset of a strongly-coupled regime.

\begin{figure}[b!]
  \includegraphics[width=.47\textwidth]{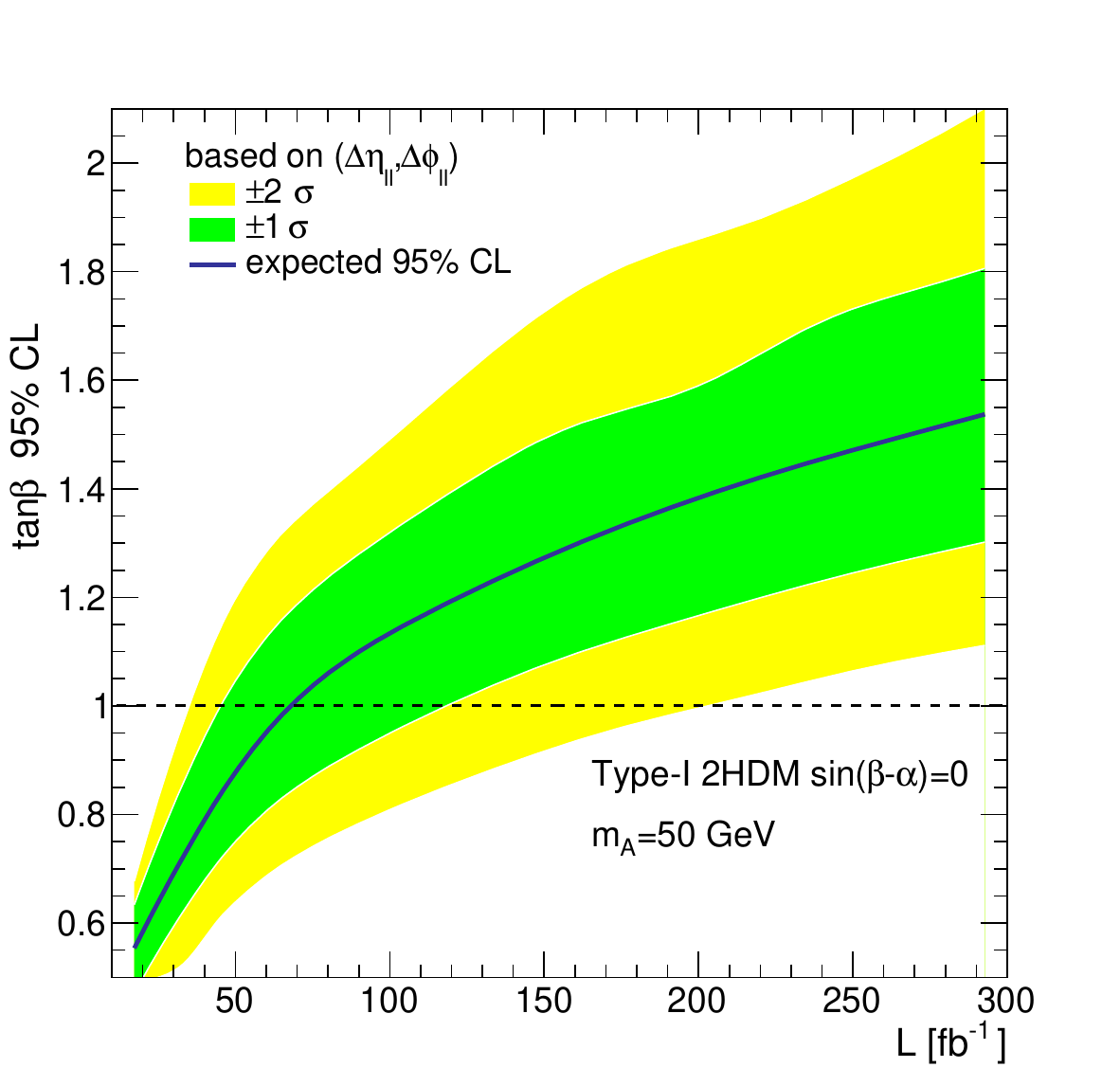}
  \caption{95\% CL bound on $\tan\beta$ as a function of the LHC luminosity from the $\ourchannel$ search.
  The binned log-likelihood analysis is based on the two-dimensional distribution $(\Delta\eta_{ll},\Delta\phi_{ll})$.
   We illustrate our results taking as exemplary case  a Type-I \textsc{2HDM} with $m_A=50$~GeV.}
  \label{fig:cls}
\end{figure}

The key feature of the \emph{alignment without decoupling} scenarios, such as those in Tab.~\ref{tab:interpretation}, is  that 
the additional Higgs states remain light or only moderately heavy. 
This means that the chief constraints on them are imposed by  
the direct LEP, Tevatron, and LHC searches. The condition 
$m_{H^{\pm}} \gtrsim 175$~GeV follows from the limits 
on charged scalars ~\cite{\searchescharged}, which hinder 
the non-standard top decay $t \to H^{\pm}b \to \tau\nu_\tau b$. 
Depending on the underlying assumptions,  
certain parameter space patches in Tab.~\ref{tab:interpretation}
with enhanced fermion Yukawas would be strongly disfavored by the 
CP-even~\cite{\searchescpeven} and CP-odd searches~\cite{\searchescpodd}. 
If the \textsc{2HDM} was to be strictly taken as an
UV completion, it would first of all be difficult to reconcile $\tan\beta < 1$ with the
$ pp \to H,A \to \gamma\gamma$ searches in the $m_{H,A} < 300$~GeV range. A more flexible approach, 
which we follow here, is to consider the \textsc{2HDM} as part of a larger UV completion, where additional 
charged states compensate the enhanced top loops in $H,A \to \gamma\gamma$. The same argument can
 be advocated to evade the indirect constraints for $\tan\beta \lesssim 1$ from charged Higgs loops in 
 $B_d-\overline{B}_d$ and $B_s^0 \to \mu^+\mu^-$~\cite{flavor}.
Also important is the role of di-tau final-states~\cite{Aad:2014vgg,Khachatryan:2014wca}.
In a strict Type-I interpretation, these would rule out a sizeable portion of the range
$\tan\beta < 1, m_A \gtrsim 105$~GeV covered by Tab.~\ref{tab:interpretation}. Such constraints
can be eluded by assuming in this case lepton-specific (viz. Type-II) couplings,
while keeping a Type-I setup for the quarks. Yet, some of the patches 
with  $\tan\beta \simeq 1$ and $m_A \simeq (110 - 180)$ GeV 
remain in tension with the recent LHC results~\cite{CMS:2015mca}. 
Likewise, 
 $\tan\beta \lesssim 0.6$ is precluded for 
 $m_{H,A} \lesssim 120$ GeV 
by the Tevatron analysis $p\bar{p} \to \tau\bar{\tau} H \to \tau\bar{\tau}\tau\bar{\tau}$~\cite{Abdallah:2004wy}.
Finally, the lowest $m_A$ edge are in conflict 
with the LHC search $pp\to t\bar{t} (H,A) \to t \bar{t} t \bar{t}$~\cite{CMS:2012ywa} and also
ruled out in part by the LEP search $e^+e^- \to hA \to b\bar{b}b\bar{b}$~\cite{Schael:2006cr}.

\begin{figure}[!t]
  \includegraphics[width=.47\textwidth]{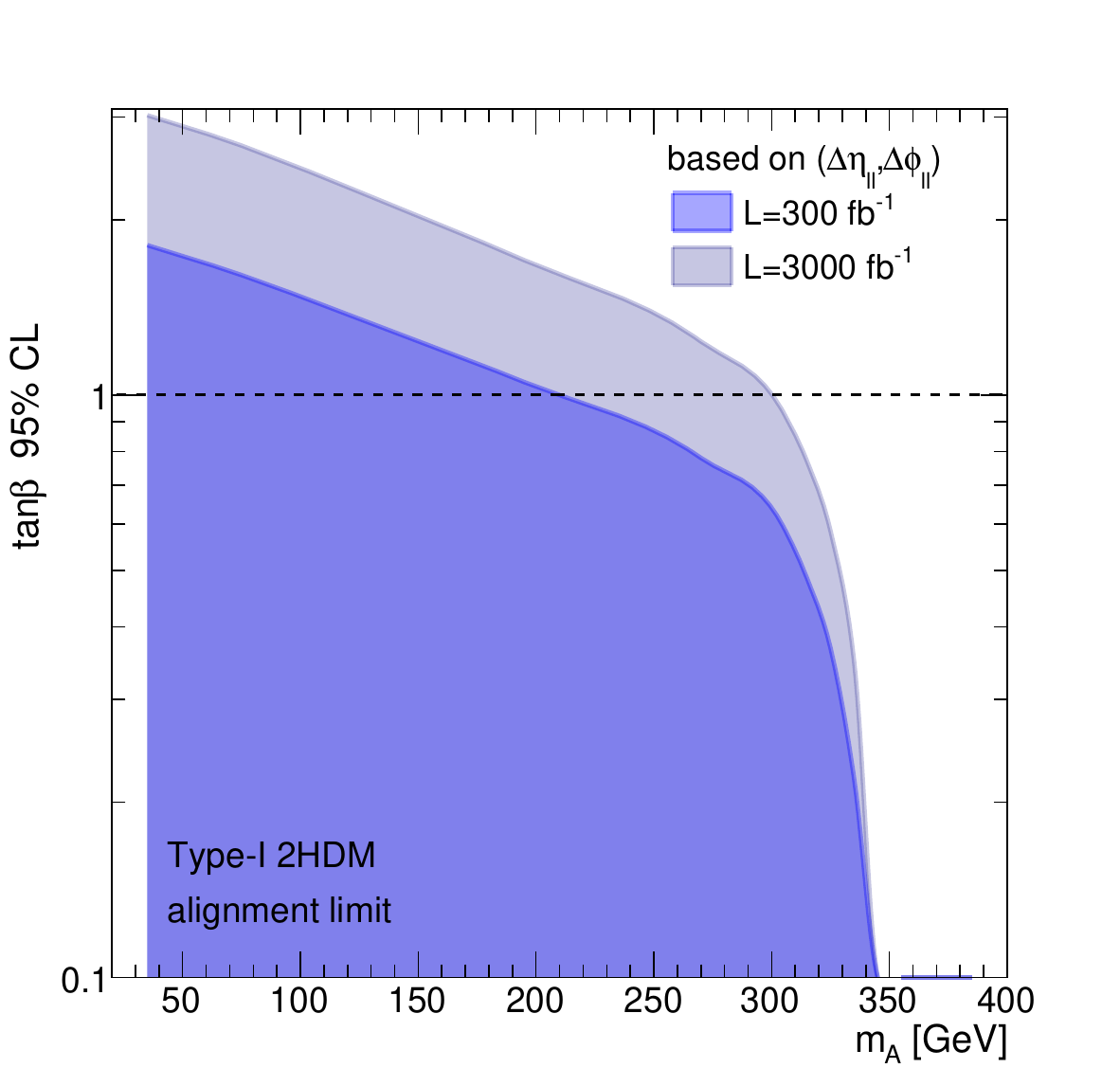}
  \caption{95\% CL exclusion region on $\tan\beta$ as a function of the  pseudoscalar mass $m_A$ for Type-I \textsc{2HDM}
   with fixed LHC integrated luminosity: ${\mathcal{L}=300~fb^{-1}}$   (blue) and ${\mathcal{L}=3000~fb^{-1}}$ (gray).}
  \label{fig:cls_mA}
\end{figure}

The benchmark regions in Tab.~\ref{tab:interpretation} serve us to promote our $\ourchannel$ search strategy 
beyond a mere Simplified Model framework, assessing its implications for a phenomenologically viable, 
UV complete, extended Higgs sector. This exercise we perform in Fig.~\ref{fig:cls}. Here we remap the 
Simplified Model constraints of Fig.~\ref{fig:cls_simp} onto the Type-I \textsc{2HDM} in the \emph{alignment limit}. 
The 95\% CL bound follows from the two-dimensional binned log-likelihood analysis of the $\bar{t}tA$ kinematics
in the variables $(\Delta\eta_{ll},\Delta\phi_{ll})$, as explained earlier on. The limits on $\tan\beta$ are obtained as
a function of the LHC integrated luminosity, for a fixed pseudoscalar mass hypothesis $m_A=50$~GeV. We find that, 
over the Run II, the LHC would be capable to bound the range $\tan\beta>1$ with only 70~$fb^{-1}$,  and climb up to
$\tan\beta>1.5$ with  300~$fb^{-1}$ of collected data. For $\tan\beta$ above these values, the number 
of signal events would not be observable. Such a remarkable sensitivity on $\tan\beta$ we can ultimately
trace back to the characteristic signal strength dependence $\mu_{\text{sig}}\propto  \cot^2\beta$ in the Type-I \textsc{2HDM},  
resulting from the  $\cot\beta$-rescaled top Yukawa along with the $\tan\beta$-independent 
$\mathcal{BR}(A\rightarrow b\bar b)$.

Additionally, in Fig.~\ref{fig:cls_mA} we display the 95\% CL bound for the same Type-I \textsc{2HDM} setup, now 
as a function of the $m_A$.  The proposed $\ourchannel$ analysis can set a limit $\tan\beta>1$ for ${20~\gev<m_{A}<210~\gev}$
with ${\mathcal{L}=300~fb^{-1}}$ and probe masses up to  ${m_A\sim 320~\gev}$ at the  high luminosity LHC ${\mathcal{L}=3000~fb^{-1}}$,
achieving a remarkable sensitivity up to $m_A\sim 2m_t$. Obviously, above the top pair production threshold, the $A\rightarrow t\bar{t}$ decay
holds the largest constraining power~\cite{Craig:2015jba}.

\section{CP measurement: $t\bar{t}A$ vs. $t\bar{t}H$}
\label{sec:cp}

In the \emph{alignment limit}, the Yukawa patterns of both the heavy
CP-even ($H$) and the CP-odd ($A$) 2HDM resonances follow,
up to sign differences, an identical dependence on $\tan\beta$, 
see Tab.~\ref{tab:2HDM}.
\begin{table}[h!]
\centering
\begin{tabular}{l | c  | c | c | c }
  \multicolumn{1}{c}{} &
  \multicolumn{2}{c|}{Type-I} &
  \multicolumn{2}{c}{Type-II} 
  \\
  \cline{2-5}
  \multicolumn{1}{c}{} &
  \multicolumn{1}{c|}{H} &
  \multicolumn{1}{c|}{A} &
  \multicolumn{1}{c|}{H} &
  \multicolumn{1}{c}{A} 
 \\
 \hline
$\kappa_{t}$ &  $-\cot\beta$ & $\cot\beta$  & $-\cot\beta$  & $\cot\beta$   \\ 
$\kappa_{b}$ & $-\cot\beta$ & $-\cot\beta$ &  $\tan\beta$& $\tan\beta$ 
\end{tabular} 
\caption{Relative coupling strength $\kappa_{t(b)}$ for top (bottom) quarks with respect to the SM 
Yukawa couplings for the  Type-I and Type-II \textsc{2HDM} scenarios in the alignment limit.}
\label{tab:2HDM}
\end{table}
This means  that both CP hypotheses would exhibit very similar signatures
when searching for heavy 2HDM resonances in the \emph{alignment limit}
through fermionic channels.   In the event of a signal excess, an immediate question 
would hence be to characterise the CP properties of the discovered heavy state. In this section, we 
demonstrate how a direct CP measurement of a heavy 2HDM candidate is possible by examining
the spin correlations in fermionic final states~\cite{Buckley:2015vsa,Ellis:2013yxa,
Boudjema:2015nda,Chang:2016mso,Demartin:2014fia}. As CP-sensitive quantity, we follow 
Ref.~\cite{Buckley:2015vsa} and choose the angular correlation variable $\Delta\phi_{ll}$: that is,
the difference in azimuthal angle around the beam axis  of  the  top-pair  leptonic decay 
products in the lab-frame. This strategy, originally applied to the 125~GeV Higgs in 
Ref.~\cite{Buckley:2015vsa}, benefits  from the more boosted $A(H)$  kinematics we 
must require for the signal to be observable. The increased CP-sensitivity in the boosted regime
is ultimately correlated  to the enhancement of the mixed helicity states $t_L\bar{t}_R+t_R\bar{t}_L$ 
in the large $p_{T, A(H)}$ region. Interestingly, this helicity state presents different  modulations in
the azimuthal top pair angle $\Delta\phi_{tt}$, which manifest as oscillations obeying $\sin\Delta\phi_{tt}$  
(resp. $\cos\Delta\phi_{tt}$) when the  top pair is produced in association with a CP-even (resp. CP-odd) 
scalar. The analytic argument is elaborated in detail in Ref.~\cite{Buckley:2015vsa}.\medskip

\begin{figure}[!t]
  \includegraphics[width=.43\textwidth]{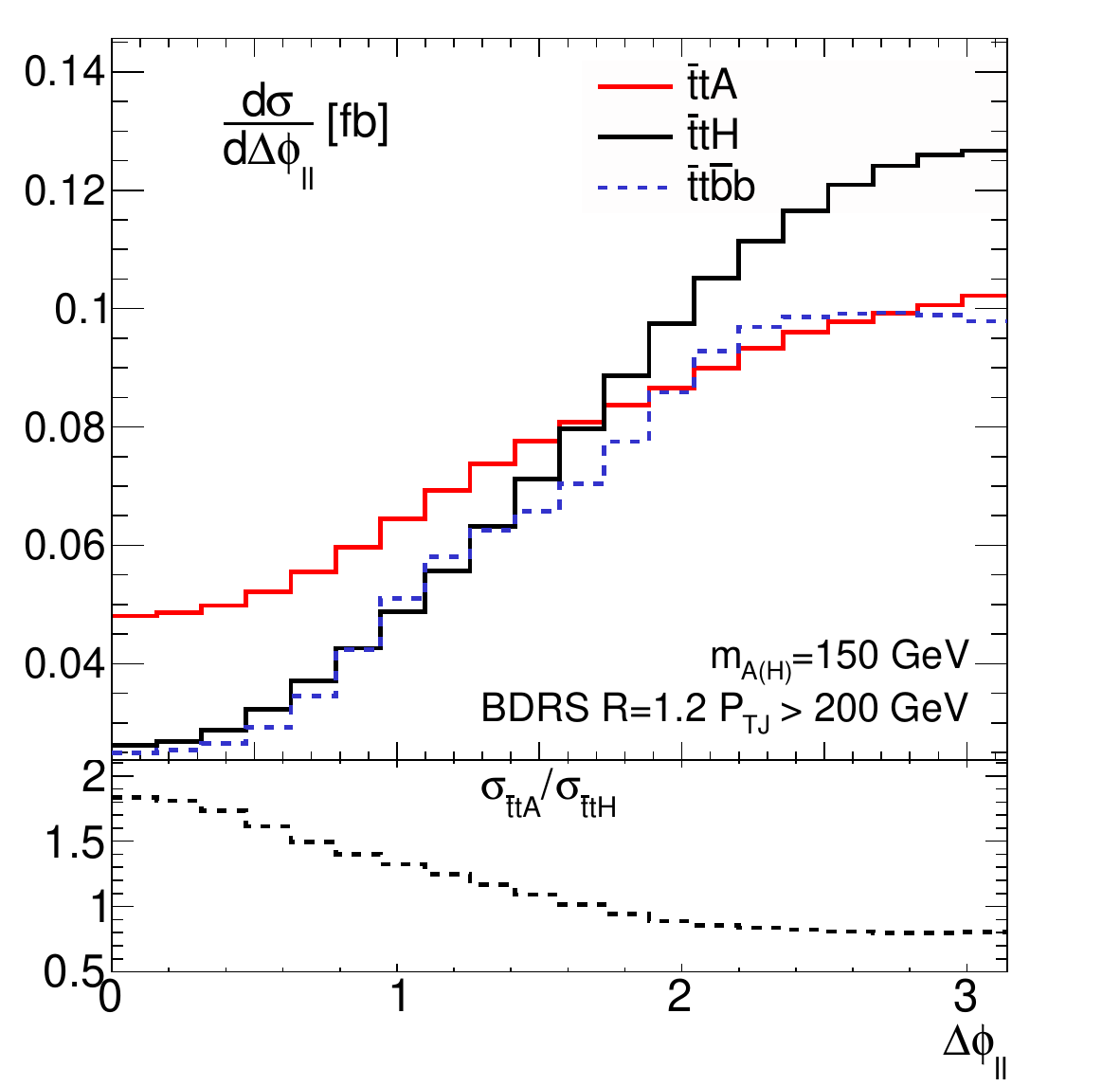}
  \caption{Azimuthal angular distribution $\Delta \phi_{ll}$ between the two leptons for the CP-odd $\bar{t}tA$ (red) 
  and CP-even $\bar{t}tH$ (black) signal hypotheses and $\bar{t}t\bar{b}b$ (blue) background. We assume a Type-I \textsc{2HDM}
  with $\tan\beta=0.5$ and $m_{A(H)}=150$~GeV.}
  \label{fig:dphill}
\end{figure}

In Fig.~\ref{fig:dphill} we compare the lab-frame $\Delta\phi_{ll}$ distributions for the CP-odd ($\bar{t}tA$) and the CP-even 
($\bar{t}tH$) signal hypotheses, together with the $\bar{t}t\bar{b}b$  background after the selection cuts given in Tab.~\ref{tab:cuts}.
The latter we supplement now with an additional cut on the di-lepton invariant mass $m_{ll}>75$~GeV. This extra requirement works as a
proxy for the di-leptonic top pair selection $m_{tt}$~\cite{Mahlon:1995zn}, further enhancing the unlike-helicity states.  To generate the 
signal events, we assume a Type-I \textsc{2HDM}  and for definiteness fix $\tan\beta=0.5$ and $m_{A(H)}=150$~GeV. As we can see, 
the analysis of the $\Delta\phi_{ll}$ distributions provides an efficient procedure to discriminate between the two competing hypotheses.
Notably, the sensitivity reaches up to  $\sigma_{\bar{t}tA}/\sigma_{\bar{t}tH}\sim2$ for small angles $\Delta\phi_{ll}\sim 0$. 

\begin{figure}[!b]
  \includegraphics[width=.45\textwidth]{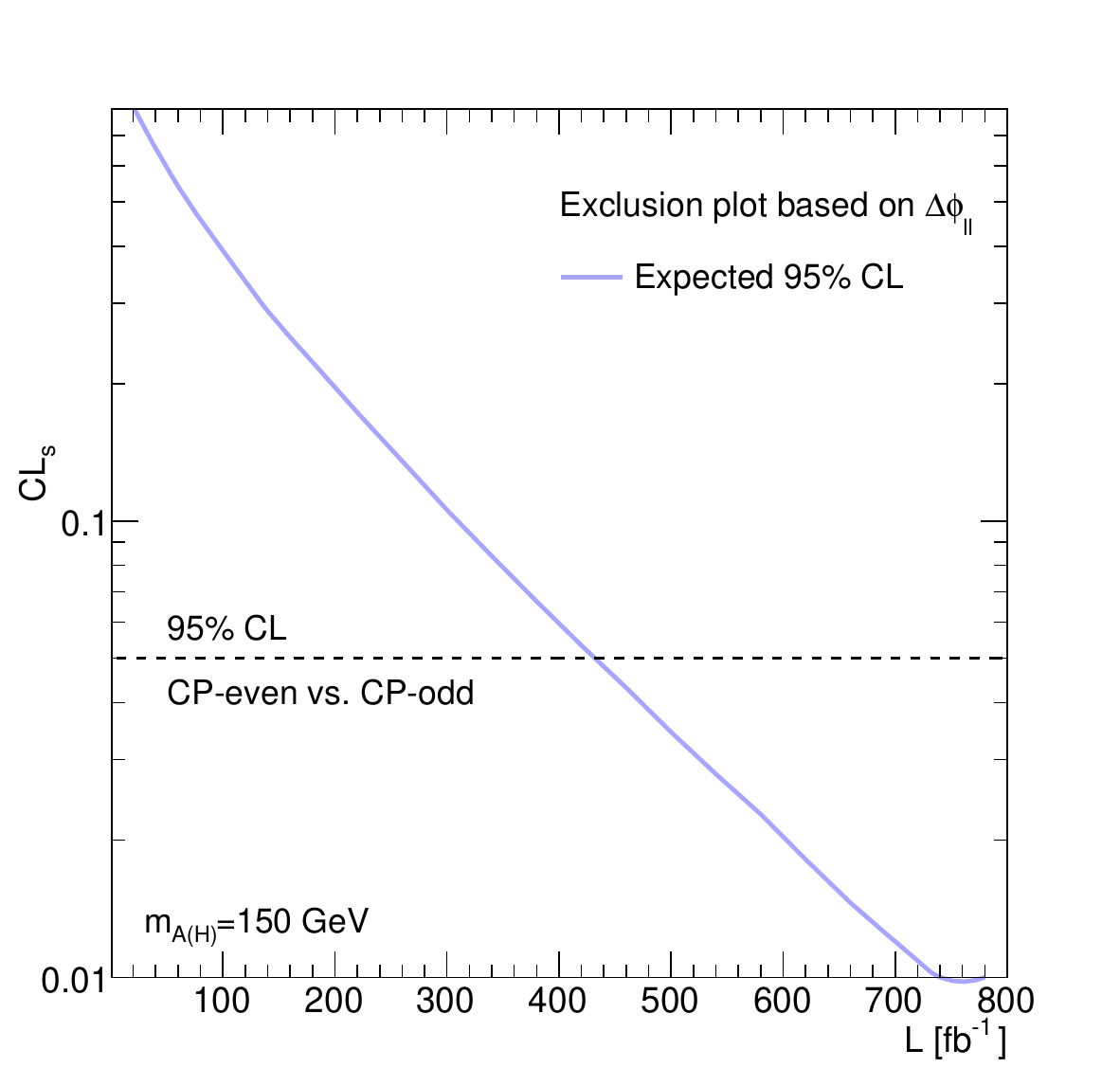}
  \caption{Luminosity needed to distinguish the CP-odd $\bar{t}tA$ from the CP-even $\bar{t}tH$ signal hypotheses at 95\% CL.
   We assume a Type-I \textsc{2HDM}  with $\tan\beta=0.5$ and ${m_{A(H)}=150}$~GeV.}
   \label{fig:cls_dphill}
\end{figure}

Besides its remarkable CP-sensitivity, the chosen variable $\Delta\phi_{ll}$ is also advantageous
from the experimental viewpoint. Thanks to the fact that it exclusively relies on the leptons, and that
it is reconstructed in the lab-frame, it is affected by rather small experimental uncertainties. In particular, 
it does not suffer from the usual uncertainties associated e.g. to the top reconstruction 
or a reference frame change.

It also worth noticing that by including the $\Delta\phi_{ll}$ distribution in the previous $\ourchannel$ analysis,
namely in the binned  log-likelihood test of Figs.~\ref{fig:cls_simp},~\ref{fig:cls} and~\ref{fig:cls_mA}, we achieve
a significantly improved signal over background ($\mathcal{S/B}$) separation. This reflects the different angular modulations 
for the $t\bar{t}A$ signal and $t\bar{t}b\bar{b}$ background that, for our showcase signal scenario in Fig.~\ref{fig:dphill}, result 
 in enhanced sensitivities ${\sigma_{\bar{t}tA}/\sigma_{\bar{t}t\bar{b}b}\sim2}$ for small  angles $\Delta\phi_{ll}\sim 0$.\medskip

Finally, in Fig.~\ref{fig:cls_dphill} we quantify the statistical power of our proposed CP discriminant,  by performing a binned 
log-likelihood test based on the $\Delta\phi_{ll}$ distribution. To focus exclusively on the power of the spin correlation
measurement, we  set both $\bar{t}tA$ and $\bar{t}tH$ event rates to the same value  $\sigma_{\bar{t}tA}$, which we compute
for our trial setup of a Type-I \textsc{2HDM}  with $\tan\beta=0.5$ and $m_{A(H)}=150$~GeV. In this plot we show the luminosity
needed to directly distinguish the two alternative CP  hypotheses at 95\%~CL in the 13 TeV LHC run. We observe that this can
be achieved after collecting $\sim 450~fb^{-1}$ of data, for the assumed number of signal events.  Let us emphasise that this 
result should be interpreted only as an upper bound. A more accurate estimate would for instance be possible by including the
three $b$-tag sample, mostly if in conjunction with possible improvements on the misstag rate~\cite{Goncalves:2015prv}, or accounting
for additional observables combined within a Boosted Decision Tree~\cite{Aad:2015gra}.

\section{Summary}
\label{sec:summary}

In this letter, we have focused on the quest for signatures of an additional pseudoscalar boson at the LHC. 
We have concentrated on moderate pseudoscalar masses and scrutinised a range of variable couplings to the
top and bottom quarks, which we describe in a Simplified Model setup, where the novel resonance interacts
with the heavy quarks through independently rescaled Yukawa couplings. To test these interactions, we devise 
a search strategy  based on the associated pseudoscalar production along with a $t\bar{t}$ pair, followed by the 
decay  $A\rightarrow b\bar{b}$ and with di-leptonic top signatures. By using jet substructure techniques, conveniently
tailored to different pseudoscalar mass ranges,  we are able to provide an ample coverage of the possible new 
resonance masses. First,  we apply our collider analysis to obtain limits on the parameter space of the 
Simplified Model. 
We then reinterpret these results in light of a ~\textsc{2HDM}
in the \emph{alignment without decoupling} limit. 
In particular we show 
that, for a Type-I \textsc{2HDM} pattern of heavy quark Yukawas, and after collecting $300$~fb$^{-1}$ of data,
it would the possible to $i$) constrain the region $\tan\beta>1.5$ at 95 $\%$ CL for a light pseudoscalar of $m_A = 50$ GeV; 
and $ii$) exclude the entire mass range $20~\gev<m_A<210~\gev$ with $\tan\beta \leq 1$. Finally, we 
analyse the spin correlations of the final-state fermions. 
We show that the difference in azimuthal angle between the leptons from
the top decays $\Delta \phi_{ll}$ in the lab-frame critically depends 
on the CP nature of the heavy 2HDM scalar, providing a direct handle on the quantum numbers of the resonance candidate. 

In view of its paramount implications for the particle physics puzzle,  unveiling footprints of additional scalars ranks very high in the wish-list
of new physics chasers. As we have demonstrated, collider searches based on heavy-quark-rich final states are called to play 
a decisive role in this task - now that the unleashed discovery power of the LHC at Run II starts crossing the borders of the 
multi-TeV territory.

\bigskip{}
\begin{center} \textbf{Acknowledgments} \end{center}

  DLV is funded by the F.R.S.-FNRS \emph{Fonds de la Recherche
  Scientifique} (Belgium). DG  is  thankful  to  the Mainz  Institute  for  
  Theoretical  Physics  (MITP)  for  its hospitality and partial support 
  during the completion of this work.


\end{document}